\documentclass[prb,twocolumn,reprint,showpacs,amsmath,amssymb,superscriptaddress,bibnotes,floatfix]{revtex4-2}
\usepackage[dvipsnames]{xcolor}
\usepackage[T1]{fontenc}
\usepackage{amsmath}
\usepackage{braket}
\usepackage{simpler-wick}
\usepackage{tikz}
\usepackage{bm}
\usepackage{dsfont}

\usepackage{graphicx}
\usepackage{hyperref}
\usepackage{placeins}
\usepackage{ulem}

\renewcommand{\v}[1]{\ensuremath{\mathbf{#1}}}
\renewcommand{\d}{\mathrm{d}}

\newcommand{\cc}[1]{{#1}^{\dagger}}

\newcommand{\comment}[1]{}

\usepackage{comment}
\usepackage{cleveref}

\begin{document}

\title{Floquet Engineering Magnetism and Superconductivity in the Square-Lattice Hubbard Model}

\author{Jan-Niklas Herre}%
 \email{jan.herre@rwth-aachen.de}
 \affiliation{Institute for Theory of Statistical Physics, RWTH Aachen University}
\author{Takuya Okugawa}
\affiliation{Department of Physics, Columbia University, New York, NY 10027, USA}
\affiliation{Max Planck Institute for the Structure and Dynamics of Matter,
Center for Free Electron Laser Science, Luruper Chaussee 149, 22761 Hamburg, Germany}
\author{Ammon Fischer}
 \affiliation{Max Planck Institute for the Structure and Dynamics of Matter,
Center for Free Electron Laser Science, Luruper Chaussee 149, 22761 Hamburg, Germany}
\author{Christoph Karrasch}
 \affiliation{Technische Universität Braunschweig, Institut für Mathematische Physik, Mendelssohnstraße 3, 38106 Braunschweig, Germany}
\author{Dante M.~Kennes}
 \affiliation{Institute for Theory of Statistical Physics, RWTH Aachen University}
 \affiliation{Max Planck Institute for the Structure and Dynamics of Matter,
Center for Free Electron Laser Science, Luruper Chaussee 149, 22761 Hamburg, Germany}
\date{\today}

\begin{abstract}
We study the interplay of magnetic order and superconductivity in the square-lattice Hubbard model under periodic driving with circularly polarized light.
Formulating diagrammatic techniques based on the random-phase approximation in terms of Floquet Green's functions, allows us to analyze fluctuation-driven unconventional pairing for weak-to-moderate interactions.
The interplay of repulsive interactions and photo-assisted hopping of electrons gives rise to a rich magnetic phase diagram featuring an antiferromagnetic-to-ferromagnetic transition prior to a Floquet Lifshitz transition. 
Close to the antiferromagnetic transition, topological $d+id$-wave superconductivity prevails in the phase diagram for a wide range of drive parameters. At intermediate-to-high frequency driving near the Floquet Lifshitz transition, superconducting orders are tuned from spin-singlet $d$-wave to spin-triplet $p$-wave character, providing an effective protocol for Floquet engineering topological superconductivity.
\end{abstract}

\maketitle
The ability to manipulate quantum materials using tailored laser fields has opened new directions for dynamically controlling their properties~\cite{Torre_2021}. In particular, Floquet engineering—the use of periodic driving to modify the effective low-energy Hamiltonian—has emerged as a promising route to realize exotic nonequilibrium states, ranging from topological phases~\cite{Oka_2009, Oka_2019, Sato_2019, Topp_2019, Wang_2013, Usaj_2014, Merboldt_2025, Rodriguez_2021} and magnetic order~\cite{Mentink_2015, Golez_2019, Sarkar_2024, Quito_2021} to light-induced superconductivity~\cite{Sentef_2016, Kennes_2019, Kennes2018, Decker_2020, Takahashi_2025, Sheikhan_2020, Kennes_2019, Hart_2019, Wang_2021, Kumar_2021, Dehghani_2021, Li_2017, Ning_2024, Sheikhan_2022, Tsuji_2015, Dasari_2018, Sentef_2017, Knap_2016, Murakami_2017, Ojeda_2021, Kuhn_2024}. Although many theoretical approaches focus on the high-frequency or low-amplitude limit, where heating is minimal and analytic control is tractable, the low-to-intermediate-frequency regime at arbitrary drive amplitude remains less understood.
Recent experiments further demonstrate that Floquet engineering of correlated materials extends beyond Mott insulators~\cite{Merboldt_2025}, underscoring the need for theoretical methods capable of addressing the weak-to-moderate coupling regime under sub-bandwidth driving, with direct relevance to Moiré-based systems~\cite{Rodriguez_2021}. 

We explore the periodically driven square-lattice Hubbard model, where previous works have investigated the possibility of Floquet engineering magnetic order in the strongly interacting limit~\cite{Mentink_2015, Quito_2021} and, for weak coupling, suppression of antiferromagnetic (AFM) order at low drive frequencies~\cite{Walldorf_2019}. 
In the limit of strong interactions, there has also been progress on the Floquet engineering of chiral $d$-wave superconductivity~\cite{Sheikhan_2022,Anan_2024}.
Here, we complement these results using the nonequilibrium Floquet Green’s function formalism \cite{Tsuji2008, Rentrop2014,Eissing_2016, Ono2018, Ono2019} that can be systematically extended into the low- to intermediate-frequency regime beyond the reach of analytical expansions. 
To assess the emergence of collective charge, spin and superconducting fluctuations, we resort to diagrammatic expansions on the level of the random-phase approximation. 
This enables us to study the impact of driven fermiology on the formation of electronic order. 
We find that, in the $t$-$t'$ Hubbard model away from half-filling driven by circularly polarized light at sub-bandwidth drive frequency, photo-assisted processes lead to a switching between AFM and ferromagnetic (FM) orders. This is fundamentally different from the Floquet Lifshitz transition effects expected in the high-frequency limit, where next-to-nearest neighbor hopping can be tuned to dominate over nearest neighbor hopping.
Close to the Floquet engineered magnetic instabilities, we find significant tunability of chiral $d/p$-wave superconductivity throughout the sub-bandwidth driving regime.

\paragraph*{Model.}\label{section: model}
We consider the square-lattice Hubbard model driven by an in-plane electric field and coupled to a metallic reservoir to facilitate a nonequilibrium steady state. The time-dependent Hamiltonian is given by
\begin{equation}
    H(t) = -\sum_{ij,\sigma } \tilde{t}_{ij}e^{-i\mathbf{A}(t)\cdot \mathbf{R}_{ij}}\cc{\hat c}_{i\sigma} \hat c_{j\sigma}  + U\sum_{i} \hat n_{i \uparrow}\hat n_{i \downarrow} + H_{\mathrm{res}} \,,
\label{eq:spinless_fermion_nn_interacting_model}
\end{equation}
where $\cc{\hat c}_{i\sigma}$ ($\hat c_{i\sigma}$) is the creation (annihilation) operator of a fermion with spin $\sigma$ on site $i$ and $\hat n_{i \sigma} = \cc{\hat c}_{i\sigma}\hat c_{i\sigma}$. $U$ and $\tilde{t}_{ij}$ are the on-site repulsion and hopping between site $i$ and $j$, respectively. For the latter, we allow for nearest-neighbor hopping $\tilde{t}$ as well as next-nearest-neighbor hopping $\tilde{t}'$ to break the perfect nesting condition and allow for superconducting order close to the magnetic transition~\cite{Romer_2015}. Circularly polarized light, breaking time-reversal symmetry, is included via Peierls substitution~\cite{Jauho_1984} with its vector potential
\begin{align}
\mathbf{A}(t) &= \frac{E_0}{\Omega}\left(\cos\Omega t, \sin\Omega t\right)\,,
\end{align}
where $E_0$ is the electrical field strength and $\Omega$ the drive frequency (drive period $T = 2\pi/\Omega$). The reservoirs are modeled by non-interacting tight-binding chains that are weakly coupled via a constant hopping term $V$ to each lattice site independently,
\begin{align}
    H_{\mathrm{res}} = \sum_{\v k_r} \varepsilon_{\v k_r}\cc{\hat d}_{\v k_r}\hat d_{\v k_r}  + \sum_{\v k \v k_r \sigma} V (\cc{\hat c}_{\v k\sigma}\hat d_{\v k_r} + \cc{\hat d}_{\v k_r} \hat c_{\v k \sigma})\;.
\end{align}
In the wide-band limit \cite{Jakobs_2010}, this is modeled as a constant hybridization $\gamma=\pi\rho_0 |V|^2$ with $\rho_0$ being the reservoir density of states. The (thermal) reservoir is defined by the reservoir distribution function $f(\omega) = (e^{(\omega-\mu)/T_{\mathrm{res}}} + 1)^{-1}$ at temperature $T_{\mathrm{res}}$ and chemical potential $\mu$. The hybridization enters as an imaginary selfenergy through the Dyson equation into the reservoir dressed Green's function of the system.\\
The system is initially in the uncorrelated normal state (paramagnetic), and we investigate how driving affects collective fluctuations and the emergence of magnetic and superconducting order in the weak-to-moderately interacting regime.
The weak coupling to the bath facilitates the formation of a Floquet steady-state that is synchronized with the drive (on the timescale $t^*\sim \gamma^{-1}$)~\cite{Tsuji_2009}. In this formalism, dissipation and heating balance each other and no other heating channels are available.
\paragraph*{High-frequency limit.}
In the limit of high driving frequency we can gain analytical insight via the Magnus expansion \cite{Magnus_1954}, which expands in $1/\Omega$ to find an effectively static Floquet Hamiltonian. The lowest order result for the Hubbard model is known \cite{Weidinger_2017, Kennes2018, Kennes_2019, Kalthoff_2019}, and we can define effective hopping parameters $\tilde{t}_{\mathrm{eff}} \approx \tilde{t}\mathcal{J}_0(E_0/\Omega)$ and $\tilde{t}'_{\mathrm{eff}} \approx \tilde{t}'\mathcal{J}_0(\sqrt{2}E_0/\Omega)$. In this limit, we can find two characteristic points. First, a Floquet Lifshitz transition \cite{Iorsh_2017, Iorsh_2024, Mohan_2018, Beaulieu_2021} at $E_0/\Omega\approx2.28$, where $\tilde{t}_{\mathrm{eff}} = \tilde{t}'_{\mathrm{eff}}$, which leads to a topological change in the Fermi surface from a continuous one to separated pockets. Second, close to the Floquet Lifshitz transition lies the dynamical localization point  \cite{Kawakami_2018}, where $\tilde{t}_{\mathrm{eff}}\to0$, at $E_0/\Omega\approx2.4$. Here, the effective model is a square lattice model on an enlarged unit cell with hopping $\tilde{t}'_{\mathrm{eff}}$ and tilted by $\pi/4$.

\begin{figure*}[t!]  
\centering
\includegraphics[width=\textwidth]{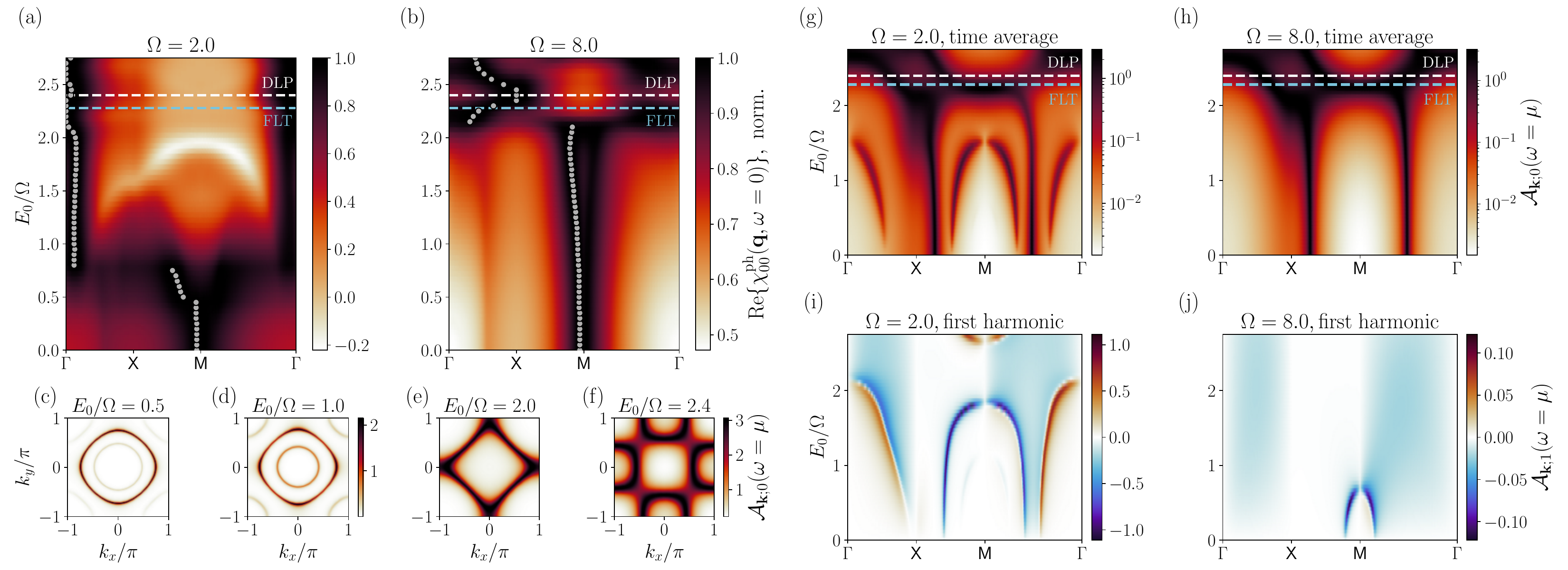}
\caption{Time averaged driven non-interacting susceptibility at $\omega=0$ and spectral function at $\omega=\mu$. Panels (a)-(b): Normalized particle-hole susceptibility in momentum space plotted over drive amplitude. (a) shows low drive frequency, (b) high drive frequency. Gray dots mark the maximum of the susceptibility for a given amplitude. The blue dashed lines mark the Floquet Lifshitz transition (FLT) and the black dashed lines mark the dynamical localization point (DLP). Panels (c), (d) show the Fermi surfaces for $\Omega=2.0$ at the transition point of the susceptibility maximum. Panels (e), (f) show the same for the high-frequency case ($\Omega=8.0$, at-bandwidth driving). Panels (g), (h) show the time-averaged spectral function at $\omega=\mu$ in momentum space over drive amplitude for both low and high frequency. Again, the dashed lines mark the Floquet-Lifshitz transition and the dynamical localization point. Panels (i), (j) show the first harmonic ($n=1$ Floquet component) of the spectral function.}
\label{fig:response_transition}
\end{figure*}

\paragraph*{Floquet random phase approximation.}
Based on the Floquet Green's function formalism \cite{Floquet_1883, Kamenev_2011, Tsuji2008,Rentrop2014,Eissing_2016, Ono2018,Ono2019}, we start from the retarded non-interacting $\mathrm{SU}(2)$-symmetric Floquet particle-hole susceptibility 

\begin{align}
    \chi^{\mathrm{ph}}_{nn'}(Q) =\frac{i}{N} \sum_{l,\v k}\int \frac{\d\omega}{2\pi}\; &G^{\mathrm{R}}_{n,n'+l}(K)G^{<}_{l,0}(K-Q) \nonumber\\
    &+ G^{<}_{n,n'+l}(K)G^{\mathrm{A}}_{l,0}(K-Q)\;, \label{eq:ph susceptibility}
\end{align}
introducing the multi-index $K=(\omega,\v k)$ for the real frequency $\omega$ and the momentum $\v k$ as well as discrete Floquet indices $n,n',l$. The Keldysh Green's functions that fulfill $G_{\v k}^{X}(t,t') = G_{\v k}^X(t+T,t'+T)$ for $X\in\{\mathrm R,\mathrm A,< \}$ were Fourier transformed twice into
\begin{equation}
    G^{X}_{K, nn'}=\frac{1}{T}\int_{0}^{T}\d\overline{t} \int_{-\infty}^{\infty}\d\tau e^{i(\omega + n\Omega) t -i(\omega + n'\Omega) t'}G_{\v k}^{X}(t, t') \;,
\end{equation}
with $\overline{t}=\frac{t+t'}{2}$ and $\tau=t-t'$ (see Supplemental Material (SM) for details~\cite{SM}).
Eq.~\eqref{eq:ph susceptibility} is a matrix in Floquet space $\hat\chi^{\mathrm{ph}}_{Q}$ and we can define the spin (s) and charge (c) susceptibility as a geometric sum within the random phase approximation
\begin{equation}
    \hat{\chi}^{\mathrm{s/c}}(Q) = \left[ \mathds{1} \mp \hat{V}_0\hat{\chi}^{\mathrm{ph}}(Q) \right]^{-1} \hat{\chi}^{\mathrm{ph}}(Q)\;,\label{eq: RPA susceptibility}
\end{equation}
where the bare (and undriven) interaction is introduced as $\hat{V}_0=U\mathds{1}$ with $\mathds{1}$ the identity matrix.
The geometric series of the spin susceptibility diverges at a critical interaction strength $U_c=1/\lambda_m$ where $\lambda_m$ is the eigenvalue with maximal real part of all matrices $\hat{\chi}^{\mathrm{ph}}(\v q, \omega=0)$, which is the Stoner criterion for the onset of magnetic order \cite{Stoner_1938}.
As long as $U<U_{\mathrm{c}}$, the system remains paramagnetic and subsidiary superconducting order driven by spin-fluctuation exchange~\cite{Schrieffer_1966, Bickers_1989,Scalapino_2012,Linder_2019, Fischer_2025} can arise, captured by the singlet (s) and triplet (t) components of the retarded effective pairing vertex
\begin{align}
 \hat{V}^{\mathrm{P,(s)}}(Q) &= \hat{V}_0 + \frac{3}{2} \hat{V}^2_0\hat{\chi}^{\mathrm{s}}(Q) - \frac{1}{2} \hat{V}^2_0\hat{\chi}^{\mathrm{c}}(Q)\;, \label{eq:singlet vertex} \\
 \hat{V}^{\mathrm{P,(t)}}(Q) &= - \frac{1}{2} \hat{V}^2_0\hat{\chi}^{\mathrm{s}}(Q) - \frac{1}{2} \hat{V}^2_0\hat{\chi}^{\mathrm{c}}(Q)\;.\label{eq:triplet vertex}
\end{align}
Unconventional superconducting order facilitated by nearby magnetic order is analyzed by 
diagonalizing a linearized gap equation using a static approximation of the vertex $\hat V^{\mathrm{P,(s)/(t)}}_{\v{k}-\v{k}'} \equiv \hat V^{\mathrm{P,(s)/(t)}}(\v{k}-\v{k}',\omega=0)$. 
The superconducting gap equation in the Floquet-Keldysh formalism is given by
    \begin{align}
     \Delta^{\mathrm{(s)/(t)}}_{\v{k};n} = \frac{i}{N}\sum_{\v{k}';l} V^{\mathrm{P,(s)/(t)}}_{\v{k}-\v{k}';n-l}\int \frac{\d \omega'}{2\pi} \mathcal{F}^{<}_{\v  k';l}(\omega')\;. \label{eq: Floquet gap time diagonal}
\end{align}
Here, we transformed the Floquet matrices into the single index notation from Ref.~\cite{Tsuji2008}.
In linear approximation, the anomalous Green's function is obtained as
\begin{align}
         \mathcal{F}^{<}_{\v k;nn'}(\omega) =&\sum_{lm}  G^{\mathrm{R}}_{\v k;n,m}(\omega) \Delta^{\mathrm{(s)/(t)}}_{\v k;m-l} G^{\mathrm >}_{-\v k;n',-l}(-\omega) \nonumber\\
    &+ G^{\mathrm{<}}_{\v k;n,m}(\omega)\Delta^{\mathrm{(s)/(t)}}_{\v k;m-l} G^{\mathrm R}_{-\v k;n',-l}(-\omega)\;.\label{eq: linearized anomalous Green Floquet DI}
\end{align}
With Eq.~\eqref{eq: linearized anomalous Green Floquet DI} we can solve Eq.~\eqref{eq: Floquet gap time diagonal} as a non-hermitian eigenvalue problem. 
An eigenvalue $\lambda_{\mathrm{sc}} > 1$ signals the onset of superconducting order, and the corresponding eigenvector contains information about the symmetry of the gap function.
Detailed derivations are provided in the SM~\cite{SM}.\\ 
In the numerical results, the units are set by the hopping $\tilde{t}=1$, leading to a bandwidth $W\sim 8$. We choose a small bath hybridization $\gamma=~0.1$, set next-to-nearest neighbor hopping to $\tilde{t}'=-0.2$ and fix the time-averaged filling of the non-interacting (but open and driven) system to $n_f=0.6$, away from half-filling. This is done through the chemical potential of the reservoir $\mu$, where the reservoir temperature is set to $T_{\mathrm{res}}=0$. We consider a drive frequency regime within the bandwidth $\Omega\leq W$, from low- ($\Omega=2.0$) to intermediate- ($\Omega=5.0$) to high-frequency driving ($\Omega=8.0$) where photo-assisted processes become relevant.

\paragraph*{Floquet induced response transition.}
In Fig.~\ref{fig:response_transition}, we present the Floquet engineered non-interacting susceptibility and spectral function. For reference, the $\tilde{t}'=0$ case has been studied in Ref.~\cite{Ono2018}.
For $t'\neq0$, the periodic drive leads to a Floquet Lifshitz transition of the time-averaged Fermi surface (given by the spectral function $\mathcal{A}_{\v k;0}(\omega=\mu)$) in the vicinity of the dynamical localization point (Fig.~\ref{fig:response_transition}(e)-(h)), which also affects the two-particle response function. Figs.~\ref{fig:response_transition}(a)-(b) show the static ($\omega=0$ but periodically driven $\Omega \neq 0$) non-interacting susceptibility evaluated 
along a path connecting the high symmetry points of the Brillouin zone 
as a function of drive amplitude-to-frequency ratio, $E_0/\Omega$. Panel (b) presents the result for the high-frequency driving, $\Omega=8.0$ where the effects of the Floquet Lifshitz transition and dynamical localization reflect as a transition of the response peak away from the $M$-point towards the $X$-point. Panel (a) shows low-frequency driving, $\Omega=2.0$. Here, the transition occurs at substantially smaller $E_0/\Omega\sim 0.8$. Before this transition, incommensurate peaks emerge at approximately $E_0/\Omega\sim 0.4$, indicated by a small shift of the peak away from the $M$-point.  
This behavior cannot be explained by either the Floquet Lifshitz transition or dynamical localization, as in the $\Omega=8.0$ case. Instead, one needs to take into account scattering processes not only within the ``main'' Fermi surface in the time-averaged component (Fig.~\ref{fig:response_transition}(c)-(d) show no significant change), but also between the Floquet side bands in the time average and higher harmonics that include photo-assisted hoppings and lead to a non-thermal electron occupation; see panels (g)-(j) and Ref.~\cite{Ono2018}. In this way, additional overlap of spectral weight with almost zero momentum transfer is generated by the drive, the process only being enabled by the scattering with multiple photons.

\begin{figure*}[t!]
\includegraphics[width=1.0\linewidth]{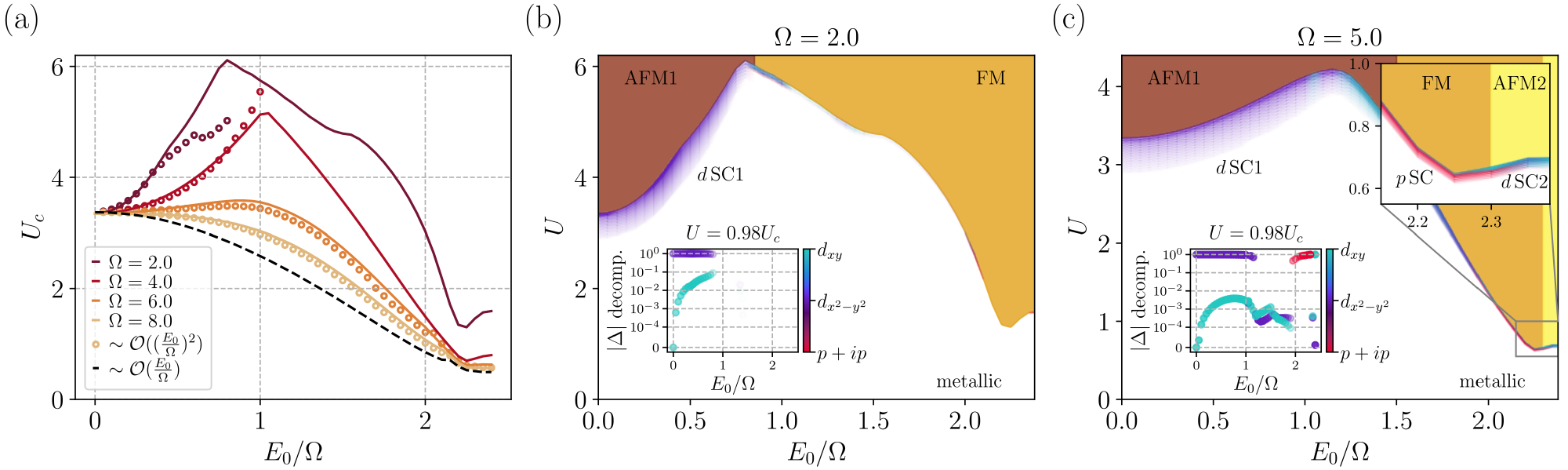}
\caption{Interplay of magnetic order and superconductivity in the weak-to-moderate interacting regime. (a) The critical interaction strength for the onset of magnetic order under circularly polarized light as a function of $E_0/\Omega$. Empty dots show results obtained by an expansion of $\chi^{\mathrm{ph}}_{00}(\v q=\v{q}_C, \omega=0)$ up to second order $\mathcal{O}((E_0/\Omega)^2)$  (see SM for details~\cite{SM}). Dashed line shows the first order expansion ($\mathcal{O}(E_0/\Omega)$) for $\Omega=8.0$. (b)-(c) Tentative nonequilibrium phase diagram of the time averaged Floquet steady state at low and intermediate-frequency driving. AFM and FM order are identified by the position of the peak of the susceptibility in momentum space. AFM1 is conventional Néel-type AFM order on the square lattice, while AFM2 also entails spin anti-alignment but on the enlarged and tilted lattice given by the next-nearest neighbors (the susceptibility diverges at the $X$-point), leading to stripe ordering. The superconducting regimes $d\mathrm{SC}1$ and $d\mathrm{SC}2$ correspond to the respective AFM transitions and show different mixing of $d_{x^2-y^2}$ and $d_{xy}$-symmetry components. Triplet $p+ip$-wave superconductivity ($p\mathrm{SC}$) is found close to the FM transition at intermediate drive frequency (panel (c)). Insets: Decomposition of the respective symmetry components of the superconducting order.}
\label{fig:phase_diagram}
\end{figure*}

\paragraph*{Floquet engineering magnetic order.}

The critical interaction strength $U_c$, above which the system develops magnetic order, is shown in Fig.~\ref{fig:phase_diagram}(a) as a function of $E_0/\Omega$. 
For the high-frequency case, $\Omega=8.0$, we observe a monotonic decrease of $U_c$, whose behavior closely follows the results obtained from a first-order expansion of $\chi^{\mathrm{ph}}_{00}(\v q=\v{q}_C, \omega=0)$, denoted by $\mathcal O(E_0/\Omega)$, where $\v{q}_C$ is the momentum giving $U_c$ in the Brillouin zone (see SM for details~\cite{SM}). This can be interpreted as the peak of the non-interacting susceptibility around the $M$-point increasing in magnitude as $\tilde{t}_{\mathrm{eff}}\to 0$ without a change in position, as shown in Fig.~\ref{fig:response_transition}(b) up to $E_0/\Omega\sim 2$. A further increase of $E_0/\Omega$ continues to decrease $U_c$, despite the change in the peak position induced by the Floquet Lifshitz transition. At the dynamical localization point ($\tilde{t}_{\mathrm{eff}}=0$), the non-zero values of $\tilde{t}'_{\mathrm{eff}}$ and $\gamma$ prevent $U_c$ from reaching zero. 
More complex behavior is to be expected in the low-frequency regime, $\Omega=2.0,4.0$. We find a suppression of magnetic order that can also be captured by an expansion up to second order, $\mathcal O((E_0/\Omega)^2)$, as denoted by the empty dots in Fig.~\ref{fig:phase_diagram}(a). 
This is similar to the results presented in Ref.~\cite{Walldorf_2019} for the perfect nesting scenario. After the photo-assisted response transition, $U_c$ starts to decrease again and the expansion breaks down. In this regime, the strong band deformation dominates over photo-assisted processes (see Fig.~\ref{fig:response_transition}(g) at $E_0/\Omega\gtrsim1.5$), restoring dynamical localization.\\
A nonequilibrium phase diagram of the time-averaged steady state is presented in Figs.~\ref{fig:phase_diagram}(b)-(c), revealing the dominant magnetic phases above the critical interaction, $U_c$ (identified in Fig.~\ref{fig:phase_diagram}(a)) as well as possible superconducting phases in the vicinity of $U_c$ (see the next section).  
For both low and intermediate frequencies with $U>U_c$, Néel-type AFM order appears at small $E_0/\Omega$. Upon increasing $E_0/\Omega$, FM order emerges since the peak position of the susceptibility is now located near the $\Gamma$-point, as shown in Fig.~\ref{fig:response_transition}(a). At intermediate-to-high frequencies ($\Omega=5.0,8.0$), a stripe AFM ordering develops on the next-to-nearest neighbor bonds, denoted as $\mathrm{AFM}2$ in Fig.~\ref{fig:phase_diagram}(c), at the dynamical localization point. At low-frequency driving, photo-assisted processes destroy the AFM2 phase, leaving only the FM phase (Fig.~\ref{fig:phase_diagram}(b)).

\paragraph*{Superconductivity near the magnetic instability.}
After establishing the modified magnetic phase diagram of the Floquet-driven Hubbard model, we continue to investigate spin-fluctuation mediated superconductivity. 
The possible symmetries of the superconducting order parameter are given by the irreducible representations of the $C_{4v}$ point group of the underlying square lattice. In addition to the $s$-wave, the irreps are given in momentum space by $\displaystyle \Delta_{d_{x^2-y^2}}(\v k) = \cos k_x - \cos k_y$, $\displaystyle \Delta_{d_{xy}}(\v k) = \sin k_x \sin k_y$ and $\displaystyle \Delta_{p_{x(y)}}(\v k) = \sin k_{x(y)}$. By projecting the superconducting gap onto the irreps we can determine the gap symmetry \cite{Romer_2015}.
In Figs.~\ref{fig:phase_diagram}(b)-(c), one can find different superconducting regimes in the vicinity of magnetic order for the low- and intermediate-frequency case. In the absence of driving ($E_0/\Omega=0$), but with coupling to the reservoir, we find spin-fluctuation induced $d_{x^2 - y^2}$ pairing, as expected from equilibrium \cite{Schrieffer_1988, Schrieffer_1989, Eberlein_2014}.    
Upon turning on the drive, the $d_{x^2 - y^2}$ ordering persists in the vicinity of Néel-type AFM ordering ($\mathrm{AFM}1$) for both low ($\Omega=2.0$) and intermediate ($\Omega=5.0$) frequencies up to $E_0/\Omega \lesssim 1$. Although the dominant pairing symmetry remains $d_{x^2-y^2}$, driving induces a mixing with $d_{xy}$ symmetry components (see inset).   %
To investigate the mixing of the sub-components and their topological properties, we calculate the Berry curvature of the associated Bogoliubov-de-Gennes Hamiltonian using Fukui's method~\cite{Fukui_2005}.
We find that a topological $d_{x^2-y^2}+id_{xy}$ order with Chern number $C=2$ is induced by driving in the vicinity of AFM1 for both $\Omega=2.0,5.0$. 
In the low-frequency case (Fig.~\ref{fig:phase_diagram}(b)), chiral $d$-wave superconductivity is suppressed as soon as the magnetic instability switches to FM order. Although triplet superconductivity could be expected to emerge in the vicinity of the FM instability, we do not find this order for the given parameters. However, in the intermediate frequency regime (Fig.~\ref{fig:phase_diagram}(c)), triplet $p+ip$-wave superconductivity can be stabilized around $E_0/\Omega\sim 2$ close to the Floquet Lifshitz transition (see also the inset in Fig.~\ref{fig:phase_diagram}(c)). With a further increase of $E_0/\Omega$, the effect of the dynamical localization point becomes evident, where $d_{x^2-y^2}+id_{xy}$ order is stabilized again, but now with the $d_{xy}$ component being dominant ($d\mathrm{SC}2$). This phase appears in the vicinity of the magnetic instability towards the stripe AFM order ($\mathrm{AFM}2$), rather than $\mathrm{AFM}1$. All orders are found to be topologically nontrivial (see SM~\cite{SM}).

\paragraph*{Discussion.}\label{section: Discussion}
We leveraged methodological advances to theoretically describe magnetic order and spin-fluctuation-mediated superconductivity in the circularly driven square-lattice Hubbard model. Introducing the Floquet random-phase approximation approach we could study the effective pairing vertex and the resulting superconducting gap under sub-bandwidth driving.
At low drive frequency, photo-assisted hopping and the resulting non-thermal electron occupation generally suppress magnetic order, whereas, at strong drive amplitude, band deformation can still lead to a dynamical localization effect that enhances magnetic order. However, the Floquet engineered fluctuations change the ordering tendency from AFM to FM order. We then investigated the fate of superconducting order in the vicinity of the magnetic instability, finding that superconductivity can generally survive and is engineered into topological $d+id$ order that is stable over a wide range of drive parameters. While low-frequency driving with high amplitude towards the FM regime destroys any superconducting order, at intermediate-frequency driving, $p$-wave superconductivity emerges together with a second type of topological $d+id$-wave ordering in the vicinity of the Floquet Lifshitz transition. 

We have shown that analytical expansions cannot be reliably extrapolated from their respective limits, but instead the complex interplay of multi-photon processes and deformed bands needs to be fully accounted for in the weak-to-moderate interacting regime. Further improvements to the method would involve its extension to truncated-unity functional renormalization group~\cite{Profe_2022, Ge2024} 
and to treat inelastic scattering processes by including frequency and momentum degrees of freedom simultaneously. Both avenues require further advances in quantum field theory simulations, a promising idea being, for example, tensor cross interpolation \cite{Shinaoka_2022,Murray_2024,Rohshap_2025}.

Our theoretical simulations highlight novel parameter regimes that are accessible in pulsed experiments~\cite{Merboldt_2025} in which Floquet engineering allows to control the subtle interplay of magnetic and superconducting order.
This offers new pathways for understanding the nature of unconventional superconductivity in cuprates~\cite{Kennes_2019}, rhombohedral multilayer graphenes~\cite{zhou2021superconductivity,holleis2025nematicity,han2025signatures} or moiré materials~\cite{Cao_2018,xia2024unconventional,guo2024superconductivity,xia2025simulatinghightemperaturesuperconductivitymoire}.

\paragraph*{Acknowledgements.}
We acknowledge fruitful discussions with Qiyu Liu. J.~H. acknowledges funding by the Deutsche Forschungsgemeinschaft (DFG, German
Research Foundation) - 508440990 - 531215165 (Research Unit ``OPTIMAL''). T.~O. acknowledges support from the JSPS Overseas Research Fellowships. Simulations were performed with computing resources granted by RWTH Aachen University under project rwth1749 as well as the HPC system Raven and Viper at the Max Planck Computing and Data Facility. 

\paragraph*{Data availability.}
The data supporting the findings of this article are not
publicly available. The data are available from the authors
upon reasonable request.
\FloatBarrier

%

\clearpage
\onecolumngrid
\appendix
    
\begin{center}
    \textbf{\large Supplemental Material for \\ ``Floquet Engineering Magnetism and Superconductivity in the Square-Lattice Hubbard Model''}
\end{center}
\section{Floquet Green's function formalism}
Since the problem is inherently nonequilibrium, we set up Green's functions on the Keldysh contour \cite{Kamenev_2011,Jakobs_2010}
\begin{align}
    G^{\mathrm{R}}_{\v{k}\sigma}(t,t') &= -i\theta(t-t')\Big\langle \left\{c_{\v{k}\sigma}(t), \cc{c}_{\v{k}\sigma}(t')\right\}\Big\rangle\,, \\
    G^{\mathrm{A}}_{\v{k}\sigma}(t,t') &= i\theta(t'-t)\Big\langle \left\{\cc{c}_{\v{k}\sigma}(t'), c_{\v{k}\sigma}(t)\right\}\Big\rangle\,, \\
    G^{<}_{\v{k}\sigma}(t,t') &= i\Big\langle \cc{c}_{\v{k}\sigma}(t') c_{\v{k}\sigma}(t)\Big\rangle\,,\\
    G^{>}_{\v{k}\sigma}(t,t') &= -i\Big\langle c_{\v{k}\sigma}(t)\cc{c}_{\v{k}\sigma}(t') \Big\rangle\,,
\end{align}
where we use the Heaviside function $\theta(t)$ and the (anti-) commutators $\{\dots\}$ and $[\dots ]$. The expectation value $\langle \dots \rangle$ is taken with respect to the paramagnetic state, which means that the Green's functions for spin $\sigma=\uparrow,\downarrow$ are identical. The goal is now to exploit the periodicity of the drive to Fourier transform the Green's functions into frequency space and obtain a time-periodic description of the Floquet steady state, in essence, applying Floquet theory \cite{Floquet_1883} to Green's functions. The time average of this steady state allows to interpret it as the effective behavior of a system with potentially completely different characteristics compared to equilibrium. The Green's functions satisfy
\begin{equation}
    G(t,t') = G(t+T,t'+T)\;,
\end{equation}
where $T=2\pi/\Omega$ is the drive period.
A double Fourier transform allows us to write, following the definition introduced in Refs.~\cite{Ono2018,Ono2019},
\begin{align}
    G_{nn'}(\omega) &=  \frac{1}{T}\int_{0}^{T}\d \overline{t} \int_{-\infty}^{\infty}\d \tau e^{i(\omega + n\Omega) t -i(\omega + n'\Omega) t'}G(t, t')\;, \label{eq:di_floquet_Greens} \\
    G(t,t') &=  \sum_{n=-\infty}^{\infty} \int_{-\infty}^{\infty}\frac{\d \omega}{2\pi} e^{-in\Omega \overline{t} -i(\omega + \frac{n}{2}\Omega) \tau}G_{n0}(\omega) \;.
\end{align}
Here we defined the central time $\overline{t}=\frac{1}{2}(t+t')$ and the relative time $\tau=t-t'$. 
The double index is actually a non-physical redundancy, which however has the advantage to allow for matrix operations like inversion and multiplication \cite{Tsuji2008}. Due to the redundancy, the Green's function needs to be restricted to the ''fundamental domain'' $F=[-\Omega/2,\Omega/2)$ \cite{Rentrop2014}.
Additionally, we can define a ''single-index'' notation as
\begin{align}
    G_{n}(\omega) &=  \frac{1}{T}\int_{0}^{T}\d \overline{t} \int_{-\infty}^{\infty}\d \tau e^{i\omega \tau +in\Omega \overline{t}}G(t, t')\;, \label{eq:si_floquet_Greens}
\end{align}
which allows for easier numerical convolution. We can translate between the two formulations as 
\begin{align}
    G_{nn'}(\omega) &= G_{n-n'}\left(\omega + \frac{n+n'}{2}\Omega\right)\;,\label{eq:di_to_si}\\
    G_{n}(\omega) &= G_{n+n',n'}\left(\omega - \frac{2n+n'}{2}\Omega\right)\;,\label{eq:si_to_di}
\end{align}
where in the second line $n'$ has to be chosen so that it lies in the fundamental domain $(\omega - \frac{2n+n'}{2}\Omega) \in F$.
Generally, we will use the version in Eq.~\eqref{eq:di_floquet_Greens} whenever the inversion of Green's functions is required or we can exploit the Floquet multiplication rule from \cite{Tsuji2008}. We use Eq.~\eqref{eq:si_floquet_Greens} whenever numerical convolution is needed.
To make Floquet Green's functions numerically tractable, the discrete Floquet indices $n,n'\in \mathbb{Z}$ must be truncated to a sufficiently large number $n_{\mathrm{max}}$. The lower the frequency of the drive and the more Fourier coefficients necessary to model the drive protocol, the more Floquet indices need to be retained.\\
The retarded Floquet Green's function of the non-interacting system dressed with the reservoir can be written in terms of the Floquet Hamiltonian and the hybridization self-energies \begin{align}
    \Sigma^{\mathrm{R/A}}_{\mathrm{res};nn'} (\omega) &= \mp i\gamma\delta_{nn'}\;,\\ 
    \Sigma^{\mathrm{<}}_{\mathrm{res};nn'} (\omega) &= 2i\gamma f(\omega + n\Omega)\delta_{nn'} \;, \\
    \Sigma^{\mathrm{>}}_{\mathrm{res};nn'} (\omega) &= -2i\gamma(1-f(\omega + n\Omega))\delta_{nn'} \;,
\end{align} as  ~\cite{Tsuji2008}
\begin{align}
    (G^{\mathrm{R}})^{-1}_{\v{k};nn'}(\omega) &= (\omega + n\Omega + i\gamma)\delta_{nn'} - \epsilon_{\v{k};n-n'}\;, \label{eq: Floquet Green's function}
\end{align}
which is a matrix that can be inverted to obtain $G^{\mathrm{R}}_{nn'}(\omega)$. Here we introduced the reservoir distribution function $f(\omega) = (e^{(\omega-\mu)/T_{\mathrm{res}}} + 1)^{-1}$ at temperature $T_{\mathrm{res}}$ and chemical potential $\mu$. We can tune the filling of the system by adjusting the bias through the bath and setting $\mu$. The Fourier components of the noninteracting dispersion are given by
\begin{equation}
    \epsilon_{\v{k},n-n'} = \frac{1}{T}\int_0^T \d t \; \epsilon_{\v{k}}(t)e^{i(n-n')\Omega t}\;.
\end{equation}
For the case of circular light implemented via the Peierls substitution we get the Fourier components of the hoppings in real space as 
\begin{align}
    \tilde{t}_{ij}^{(n)} = \tilde{t}_{ij}\mathcal{J}_{-n}\left(\frac{E_0|\v R_{ij}|}{\Omega}\right)e^{in\theta_{ij}}\;,
\end{align}
where $\mathcal{J}_n$ denote Bessel functions of the first kind and $\v R_{ij} = |\v R_{ij}|(\cos \theta_{ij}, \sin \theta_{ij})$ \cite{Anan_2024}. The dispersion $\epsilon_n(\v k)$ follows from Fourier transforming each Floquet component into momentum space. 

The other Green's functions follow as
\begin{align}
    G^{\mathrm{A}}_{nn'}(\omega) &= \cc{(\hat G^{\mathrm{R}})}_{nn'}(\omega)\;, \\
    G^{\mathrm{</>}}_{nn'}(\omega) &= ( \hat G^{\mathrm{R}}\hat \Sigma^{\mathrm{</>}}_{\mathrm{res}}\hat G^{\mathrm{A}})_{nn'}(\omega) \;,
\end{align}
where hats denote matrices in Floquet space and $(\hat{A}\hat{B})_{mn} \equiv\sum_{l}A_{ml}B_{ln}$ is conventional matrix multiplication over Floquet indices \cite{Tsuji2008}. The Floquet spectral function is given by
$\mathcal A_{n}(\omega)=-\frac{1}{\pi}\Im G^{\mathrm R}_{n}(w)$.
\section{Floquet random phase approximation (RPA)}
 In the following, we provide an alternative but equivalent formulation of the Floquet random phase approximation given in the main text, based on channel decompositions of the nonequilibrium Bethe-Salpeter equations for
two-particle correlation functions. This serves to clarify how the fluctuation exchange approximation of the effective pairing vertex remains valid in the Floquet-Keldysh picture.
 The central components of the Bethe-Salpeter equations are the non-interacting particle-hole and particle-particle susceptibilities
\begin{align}
    \chi^{\mathrm{ph};\mathrm{R}}_{\v{q}}(t,t') &=\frac{1}{N} \sum_{\v{k}}\Pi^{\mathrm{ph;R}}_{\v{k},\v{q}}(t,t')\;,\\
    \chi^{\mathrm{pp};\mathrm{R}}_{\v{q}}(t,t') &=\frac{1}{N} \sum_{\v{k}} \Pi^{\mathrm{pp;R}}_{\v{k},\v{q}}(t,t')\;,
\end{align}
where we defined the corresponding pair propagators
\begin{align}
    \Pi^{\mathrm{ph;R}}_{\v{k},\v{q}}(t,t') &=i G^{\mathrm{R}}_{\v{k}}(t,t')G^{<}_{\v{k} - \v{q}}(t',t) + iG^{<}_{\v{k}}(t,t')G^{\mathrm{A}}_{\v{k} - \v{q}}(t',t)\;,\\
    \Pi^{\mathrm{pp;R}}_{\v{k},\v{q}}(t,t')&=iG^{\mathrm{R}}_{\v{k}}(t,t')G^{>}_{-\v{k} + \v{q}}(t,t') + iG^{<}_{\v{k}}(t,t')G^{\mathrm{R}}_{-\v{k} + \v{q}}(t,t')\;.
\end{align}
Fourier transforming the susceptibilities using Floquet Green's functions yields \cite{Ono2019}
\begin{align}
    \chi^{\mathrm{ph};\mathrm{R}}_{\v{q};nn'}(\omega) &=\frac{i}{N} \sum_{\v{k},l} \int_{-\infty}^{\infty}\frac{\d \omega'}{2\pi} G^{\mathrm{R}}_{\v{k};n,n'+l}(\omega')G^{<}_{\v{k} - \v{q};l,0}(\omega'-\omega) + G^{<}_{\v{k};n,n'+l}(\omega')G^{\mathrm{A}}_{\v{k} - \v{q};l,0}(\omega'-\omega)\;, \label{eq:ph susceptibility} \\
    \chi^{\mathrm{pp};\mathrm{R}}_{\v{q};nn'}(\omega) &=\frac{i}{N} \sum_{\v{k},l} \int_{-\infty}^{\infty}\frac{\d \omega'}{2\pi} G^{\mathrm{R}}_{\v{k};n,n'+l}(\omega')G^{>}_{-\v{k} + \v{q};l,0}(-\omega' + \omega) + G^{<}_{\v{k};n,n'+l}(\omega')G^{\mathrm{R}}_{-\v{k} + \v{q};l,0}(-\omega' + \omega)\;.\label{eq:pp susceptibility}
\end{align}
This is again a matrix in Floquet space like the single particle Green's functions. In the equilibrium limit(zero amplitude or zero frequency) one can show that the retarded susceptibilities are equivalent to the equilibrium susceptibility \cite{Ono2019}. In the following, we will drop the $\mathrm R$ label, $\hat \chi^{\mathrm{ph};\mathrm{R}}_{\v{q}}(\omega)\equiv \hat \chi^{\mathrm{ph}}_{\v{q}}(\omega)$, since we will only ever consider the retarded component explicitly.\\

The Bethe-Salpeter equations in terms of vertices $\Gamma$ (two-particle correlation functions with amputated external legs), can be decomposed into three different channels classified by their two-particle reducibility: Particle-particle (P), crossed particle-hole (C) and direct particle-hole (D) channel (on a one-loop level). The full vertex can then be written using projections between the three channels.
\begin{align}
    \Gamma^{1'2'|12} (\omega_1,\v{k}_1;\omega_2,\v{k}_2;\omega_3, \v{k}_3) = &\Lambda     + P^{-1}[P]\nonumber\\
    &+ C^{-1}[C]+ D^{-1}[D]\;,
\end{align}
where $\Lambda$ is the irreducible part of the vertex, which is commonly approximated in the Parquet approximation to be the bare vertex. The dependency on four arguments is reduced to three due to (discrete) translation invariance in space and time. The indices $1'2'|12$ are the Keldysh indices which behave exactly like orbital indices, obeying several symmetries, which in the end allows to define R, A, K components of the vertex on the RPA level \cite{Jakobs_2010}. The projectors $X^{-1}[\cdot]$ reorder the Keldysh, momentum, and frequency indices into channel-native ordering with one bosonic transfer momentum (frequency) and two fermionic momenta (frequencies):
\begin{align}
    P[\Gamma](Q_p,K_p,K'_p) &= \Gamma(Q_p, K_p + K_p', K_p' - K_p)\;,\\
    C[\Gamma](Q_c,K_c,K'_c) &= \Gamma(K_c - K_c', Q_c, K_c + K_c')\;,\\
    D[\Gamma](Q_d,K_d,K'_d) &= \Gamma(K_d + K_d', K_d - K_d', Q_d)\;,
\end{align}
where we introduced the multi-index $K := (\omega,\v{k})$ and defined the momenta in the corresponding channels as 
\begin{align}
    Q_p &= K_1 \;, \;\; K_p = \frac{1}{2}(K_2-K_3)\;, \;\;K_p' = \frac{1}{2}(K_2 + K_3)\;, \\
    Q_c &= K_2 \;, \;\; K_c = \frac{1}{2}(K_1 -K_3)\;,\;\;K_c' = \frac{1}{2}(K_1 + K_3)\;, \\
    Q_d &= K_3 \;, \;\; K_d = \frac{1}{2}(K_1 +K_2)\;,\;\;K_d' = \frac{1}{2}(K_1 -K_2)\;.
\end{align}
The projection of the Keldysh indices turns out to be trivial due to the symmetries described in Ref.~\cite{Jakobs_2010} and is thus not explicitly noted.
One finds that the projections still work if one requires periodicity in time for the interaction vertex $\Gamma(t_1,t_1';t_2,t_2') = \Gamma(t_1+T,t_1'+T;t_2+T,t_2'+T)$, however, it requires an additional Floquet index as in the single particle case \cite{Rentrop2014}, which again projects trivially between the channels. The full Bethe-Salpeter equations are of course very complicated to solve numerically, and out of scope for anything beyond quantum dots; see a benchmark of parquet and multiloop FRG in that context in Ref.~\cite{Ge2024}. Instead we follow the random phase approximation and neglect any feedback between the channels by writing fully decoupled equations for each channel, where we additionally neglect all fermionic momenta (frequencies). For clarity, we first present the real-time version of the RPA equations, and then Fourier transform twice to obtain the Floquet steady state counterpart. For each retarded channel $X \in\{\mathrm{P,C,D}\}$, it is
\begin{equation}
    X^{\mathrm{RPA}}_{\v{q}_X}(t,t') = X[\Gamma^{0}]_{\v{q}_X}(t,t') + \xi_X\int\d t_1\d t_2\;X[\Gamma^{0}]_{\v{q}_X}(t,t_1)\chi^{\xi}_{\v{q}_X}(t_1,t_2) X^{\mathrm{RPA}}_{\v{q}_X}(t_2,t') \;, \label{eq: real time RPA equations}
\end{equation}
where $\xi\in\{\text{pp},\text{ph}\}$ and $\xi_P=1/2$, $\xi_C=1$, $\xi_D=-1$ are channel specific prefactors, following notation in Ref.~\cite{Fischer_2025}. Here, we additionally used the fact that the Keldysh rotation decouples the equations in Keldysh space just like in the Dyson equation \cite{Jakobs_2010}. We denote by $\Gamma^{0;\mathrm{R}}$ the initial (bare) interaction which we generally assume to be time-independent. Restricting ourselves to the Hubbard model, it is $X[\Gamma^{0;\mathrm{R}}]_{\v{q}_X}(t,t') = \delta_{tt'}U\equiv V_0$ in each channel. 
We can Fourier transform Eq.~\eqref{eq: real time RPA equations} enabling us to turn them into matrix equations
\begin{equation}
    \hat{X}^{\mathrm{RPA}}_{\v{q}_X}(\omega_X) = \hat{V}_0 + \xi_X \hat{V_0}\hat{\chi}^{\xi}_{\v{q}_X}(\omega_X)\hat{X}^{\mathrm{RPA}}_{\v{q}_X}(\omega_X) \;. \label{eq: Floquet RPA equations}
\end{equation}
 The bare interaction is defined as $\hat{V}_0=V_0\mathds{1}$, where $\mathds{1}$ denotes the identity matrix. The solution to this equation is a geometric series. The RPA expression for the particle-hole channel is
\begin{equation}
    \hat{V}^{\mathrm{C}}_{\v{q}_C}(\omega_C) = \left(\left[ \mathds{1} - \hat{V}_0\hat{\chi}^{\mathrm{ph}}\right]^{-1} \hat{V}_0\right)_{\v{q}_C}(\omega_C) \label{eq: RPA PH vertex}
\end{equation}
The series diverges at a critical interaction strength $U_c=1/\lambda_m$ where $\lambda_m$ is the eigenvalue of all matrices $\hat{\chi}^{\mathrm{ph}}_{\v{q}_C}(\omega_C)$ with maximal real part, which is the well known Stoner criterion for the onset of magnetic order. In the following, we always require staying below $U_c$, i.e., in the paramagnetic phase.\\
To investigate how spin fluctuations near the magnetic instability might lead to an effective attractive interaction in the pairing channel, we utilize the projection method known from the fluctuation exchange approximation (FLEX) \cite{Bickers_1989, Scalapino_2012}. For interactions $U<U_c$ we can project the RPA expression for the particle-hole channel Eq.~\eqref{eq: RPA PH vertex} into the pairing channel, explicitly utilizing crossing relations between the C and D-channel
\begin{align}
    \hat{V}^{\mathrm P}_{\v{k}_P,\v{k}'_P}(\omega_P, \omega_P') 
    =& \hat{V}_{0} + \frac{\hat{V}_0^2\hat{\chi}^{\mathrm{ph}}_{\v{k}_P + \v{k}_P'}(\omega_P + \omega_P')}{\mathds{1} - \hat{V}_0\hat{\chi}^{\mathrm{ph}}_{\v{k}_P + \v{k}_P'}(\omega_P + \omega_P')}
    + \frac{\hat{V}_0^3\bm{\chi}^{\mathrm{ph}}_{\v{k}_P - \v{k}_P'}(\omega_P - \omega_P')^2}{\mathds{1} - \hat{V}_0^2\hat{\chi}^{\mathrm{ph}}_{\v{k}_P - \v{k}_P'}(\omega_P - \omega_P')^2} \;. \label{eq: FLEX pairing vertex}
\end{align}
In the second line, we explicitly enforced the SU(2)-symmetric case in the $\{\uparrow\downarrow\uparrow\downarrow\}$ sector, fully removing the implicit spin dependency. To obtain the singlet (triplet) pairing vertex, we (anti)symmetrize the vertex $\hat V^{\mathrm s (\mathrm t),\mathrm P} = \hat V^{\mathrm P}_{\v k_P,\v k'_P} \pm \hat V^{\mathrm P}_{\v k_P,-\v k'_P}$. \\
We can rewrite this to obtain the formulation of the effective pairing vertex in the main text (Eqs.~($7$), ($8$)) by enforcing the $SPOT$-symmetry condition \cite{Linder_2019} and by introducing the RPA spin and charge susceptibilities (Eq.~($6$) in the main text). 
\subsection{Superconducting gap}
\begin{figure}
\includegraphics[width=.5\linewidth]{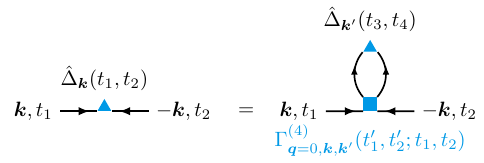}
\caption{The linearized gap equation as a general eigenvalue problem. The hat signifies $\hat\Delta$ being a matrix in Keldysh indices, while $\Gamma^{(4)}\equiv\Gamma^{1'2'|12}$ is the vertex projected into the pairing channel with all possible Keldysh indices. We only consider zero-momentum Cooper pairs, thus setting $\v q = 0$.}
\label{fig:linearized_gap_time_dependent}
\end{figure}
With an approximation for the effective pairing vertex induced by spin fluctuations at hand we can investigate the corresponding order parameter, which is the superconducting gap. In its most general form, we can write the order parameter as a Keldysh contour integral with the anomalous Green's function  
\begin{equation}
    \Delta_{\v{k}}^{12} (t_1,t_2) = \frac{i}{N}\sum_{\v{k'}}\int_{t_1',t_2'}\; V^{\mathrm{P};1'2'|12}_{\v{k},\v{k}'}(t_1',t_2';t_1,t_2)\mathcal{F}^{1'2'}_{\v{k}'}(t_1',t_2')\;, \label{eq: general real time superconducting gap}
\end{equation}
where for the Keldysh indices, a sum over repeated indices is implied. To obtain a description of the superconducting gap in the Floquet steady state, first Fourier transform Eq.~\eqref{eq: general real time superconducting gap} twice with respect to center and relative times. The general version of the vertex that is dependent on four times can be transformed utilizing the definition in Ref.~\cite{Rentrop2014}, called ''single index'' notation therein. Using the native frequency definition of the pairing channel and setting the bosonic transfer momentum to zero, we arrive at
    \begin{align}
     \Delta^{12}_{\v{k};n}(\omega) = \frac{i}{N}\sum_{\v{k}';l} \int \frac{\d \omega'}{2\pi} V^{\mathrm P; 1'2'|12}_{\v{k},\v{k}';n-l}(-(n+l)\Omega,\omega,\omega')\mathcal{F}^{1'2'}_{\v  k';l}(\omega')\;. \label{eq: Floquet gap general}
\end{align}
We concentrate on the gap at equal time $t\equiv t_1=t_2$, ($\omega=0$), which has only one independent component on the Keldysh contour \cite{Ge2024, Jakobs_2010}. 
It is given in terms of the lesser anomalous Green's function as 
\begin{equation}
    \Delta_{\v{k}} (t) = \frac{i}{N}\sum_{\v{k}'} V^{\mathrm{P};\mathrm{R}}_{\v{k},\v{k}'}(t)\mathcal{F}^{<}_{\v{k}'}(t,t)\;, \label{eq: real time superconducting gap}
\end{equation}
where, explicitly,
\begin{equation}
    \mathcal{F}^{<}_{\v{k}}(t,t')  = -i \langle \hat c_{\bm{k}\uparrow} (t') \hat c_{-\bm{k}\downarrow} (t) \rangle \;.
\end{equation}
Further, we evaluate the vertex at zero frequency (static approximation), which for the single (triplet) gap leads to
    \begin{align}
     \Delta^{(s)/(t)}_{\v{k};n} = \frac{i}{N}\sum_{\v{k}';l} V^{\mathrm P,(s)/(t)}_{\v{k}-\v{k}';n-l}\int \frac{\d \omega'}{2\pi} \mathcal{F}^{<}_{\v  k';l}(\omega')\;, \label{eq: Floquet gap time diagonal}
\end{align}
using the singlet (triplet) pairing vertex given by Eqs.~($7$), ($8$) in the main text.
In the considered drive-parameter regime, neglecting the frequency dependence of the vertex remains a reasonable assumption (within the RPA) although finite frequency effects overall are likely to reduce the pairing strength.
 The real-time version of the order parameter can be recovered as 
\begin{equation}
    \Delta_{\v k}(t) = \sum_n e^{-in\Omega t} \Delta_{\v k; n}\;.
\end{equation}
\subsection{Linearized anomalous Green's function}
Close to the transition temperature $T_c$ we can express the anomalous Green's function through the normal (bare) Green's function as 
\begin{align}
\mathcal{F}^{<}_{\v{k}}(t,t') = \int \d t_1 [ &G^{R}_{\v{k}}(t,t_1)G^{>}_{-\v{k}}(t',t_1) \nonumber\\
&+ G^{<}_{\v{k}}(t,t_1)G^{R}_{-\v{k}}(t',t_1) ]\Delta_{\v{k}}(t_1)\;, \label{eq: linearized anomalous Greens}
\end{align}
where the Keldysh summation follows from the Langreth rules similar to the pair propagators. The diagrammatic form of the general time-dependent linearized gap is shown in Fig.~\ref{fig:linearized_gap_time_dependent}. 
Fourier transforming Eq.~\eqref{eq: linearized anomalous Greens} yields in single-index notation
\begin{align}
    \mathcal{F}^{<}_{\v{k};n}(\omega) &= \sum_{l,m} \big[ G^{\mathrm{R}}_{\v k;n-l-m}(\omega + \frac{l+m}{2}\Omega)G^{\mathrm{>}}_{-\v k;l}(-\omega + \frac{n-l}{2}\Omega)+ (G^{\mathrm{R}} \longleftrightarrow G^{\mathrm{<}})\big] \Delta_{\v k;m}\;.\label{eq: linearized anomalous Green Floquet}
\end{align}
Using the double-index version of the Floquet Green's functions we have
\begin{align}
    \mathcal{F}^{<}_{\v k;nn'}(\omega) =\sum_{lm}  &G^{\mathrm{R}}_{\v k;n,m}(\omega) \Delta_{\v k;m-l} G^{\mathrm >}_{-\v k;n',-l}(-\omega) \nonumber\\&
    + G^{\mathrm{<}}_{\v k;n,m}(\omega)\Delta_{\v k;m-l} G^{\mathrm R}_{-\v k;n',-l}(-\omega)\;.\label{eq: linearized anomalous Green Floquet DI}
\end{align}
We can plug Eq.~\eqref{eq: linearized anomalous Green Floquet DI} into Eq.~\eqref{eq: Floquet gap time diagonal} and solve it as a non-Hermitian eigenvalue problem \cite{Bickers_1989, Fischer_2025}.
The onset of superconducting order is signified by the eigenvalue with maximum real part becoming larger than $1$. The symmetry of the superconducting order is encoded in the corresponding eigenvector.
\subsection{Bogoliubov-de-Gennes Hamiltonian}
Another way to compute the superconducting gap is to extend the formulation of the Hamiltonian to particle-hole (Nambu) space and consider the corresponding Bogoliubov-de-Gennes (BdG) Hamiltonian
\begin{align}
H^{\mathrm{BdG}}_{\v k}(t) = \begin{pmatrix}
   \epsilon_{\v k}(t) & \Delta_{\v k} (t) \\
    \Delta_{\v k}^* (t) & -\epsilon_{-\v k}(t)  
\end{pmatrix}\;, \label{eq: H_BdG}
\end{align}
for which we can define the Floquet Greens function in Nambu space as 
\begin{align}
\left[ \mathcal{G}^{-1}\right]^{\mathrm{R}}_{\v k;nn'}(\omega) &=\begin{pmatrix}
   \mathbf{G}_{\v k}^{\mathrm{R}}(\omega)  & \mathbf{F}^{\mathrm{R}}_{\v k}(\omega) \\
   \mathbf{\Tilde{F}}^{\mathrm{R}}_{\v k}(\omega)  & \mathbf{\Tilde{G}}_{\v k}^{\mathrm{R}}(\omega) 
\end{pmatrix}^{-1}_{nn'}\\
&= \begin{pmatrix}
   \omega_{nn'} - \epsilon_{\v k;n-n'}  & -\Delta_{\v k;n-n'} \\
    -\Delta_{\v k;n-n'}^* & \omega_{nn'}+\epsilon_{-\v k;n-n'} 
\end{pmatrix}\,,
\label{eq: G_BdG}
\end{align}
where we introduced $\omega_{nn'} = (\omega + n\Omega+ i\gamma)\delta_{nn'}$.
We extract the lesser anomalous Greens function from the corresponding matrix element in 
\begin{equation}
    \mathbf{\mathcal{G}}^{<}(\omega) = \frac{1}{2}(\mathbf{\mathcal{G}}^{A}(\omega) - \mathbf{\mathcal{G}}^{R}(\omega) +\mathbf{\mathcal{G}}^{K}(\omega))\;.
\end{equation}
We use the BdG-Hamiltonian to analyze the band structure and compute the Berry curvature and Chern number.
%
\section{Implementation}
The most numerically intensive task is to compute the non-interacting susceptibility in Eq.~\eqref{eq:ph susceptibility}.
For that we need to truncate the infinite Fourier series that builds the Floquet components to a finite amount $n_{\mathrm{max}}$ so that we deal with matrices of size $n_{\mathrm{max}}\times n_{\mathrm{max}}$.
To switch from the double to single index index variant (Eqs.~\eqref{eq:di_to_si},~\eqref{eq:si_to_di}) we "stitch" the components of the double index Green's function together, which are each only defined on the fundamental domain $F$. Here, we need to keep in mind that the truncation cuts of the accessible frequency range down to $\sim n_{\mathrm{max}}\Omega/2$. To reduce the required $n_{\mathrm{max}}$ to properly resolve the Green's function over the whole real frequency axis, we extrapolate its tails with an algebraic fit. This is possible since the main contribution of the Floquet replicas is within the bandwidth $W$. In the $\omega\to\infty$ limit the Floquet components of the Green's function follow the known algebraic decay $\sim |\omega|^{-\alpha}$. Thus, we only need enough Floquet components (qualitatively $n_{\mathrm{max}}\sim W/\Omega$) to resolve the center region within the band until the algebraic tail dominates and apply extrapolation for large $|\omega|$. In practice, we checked for convergence in $n_{\mathrm{max}}$ with extrapolation applied (a faulty extrapolation would hinder convergence). This leads to a truncation of the Fourier series down to $n_{\mathrm{max}}=9-14$ components, sufficient for drive frequencies $\Omega\gtrsim 2.0$. The frequency integral is solved using an adaptive quadrature routine using the Gauss-Kronrod 21 rule. At $T_{\mathrm{res}}=0$ it can also be computed analytically with an eigenvalue decomposition of the Floquet Green's functions as we do not feed frequency-dependent self-energies back into the computation and stay at a one-shot level. Having dealt with the Floquet and frequency dependency, the Floquet Green's function also needs to be evaluated over a sufficiently dense momentum grid that can resolve all features of the Fermi surface. The momentum sum can be solved using the convolution theorem, significantly enhancing numerical efficiency. We use a grid with $N_{\v{q},x} = N_{\v{q},y}=128$ momentum points.

Next, we Eq.~\eqref{eq: FLEX pairing vertex} to obtain the effective pairing vertex that is then plugged into Eq.~\eqref{eq: Floquet gap time diagonal} together with the lesser anomalous Greens function given by Eq.~\eqref{eq: linearized anomalous Green Floquet DI}. This can be diagonalized for the eigenvalue with the largest real part with a sparse eigenvalue solver.
For the integration of the anomalous Green's function we can employ analytical frequency integration via eigen decomposition together with the convolution theorem over momentum points, as long as we approximate the vertex to be static in frequency space.
We implement everything using efficient parallelization schemes on the GPU provided by the \textit{JAX} Python framework. The code is only memory-bound and runs in $\sim15$ minutes on one NVIDIA H100 GPU. Increasing resolution would require a multi-node GPU implementation but is in principle feasible, as well as including the frequency dependency of the vertex and a fully self-consistent FLEX simulation. This is a matter of ongoing work. 

\section{Floquet engineering the effective pairing vertex}
\begin{figure}
\includegraphics[width=1\linewidth]{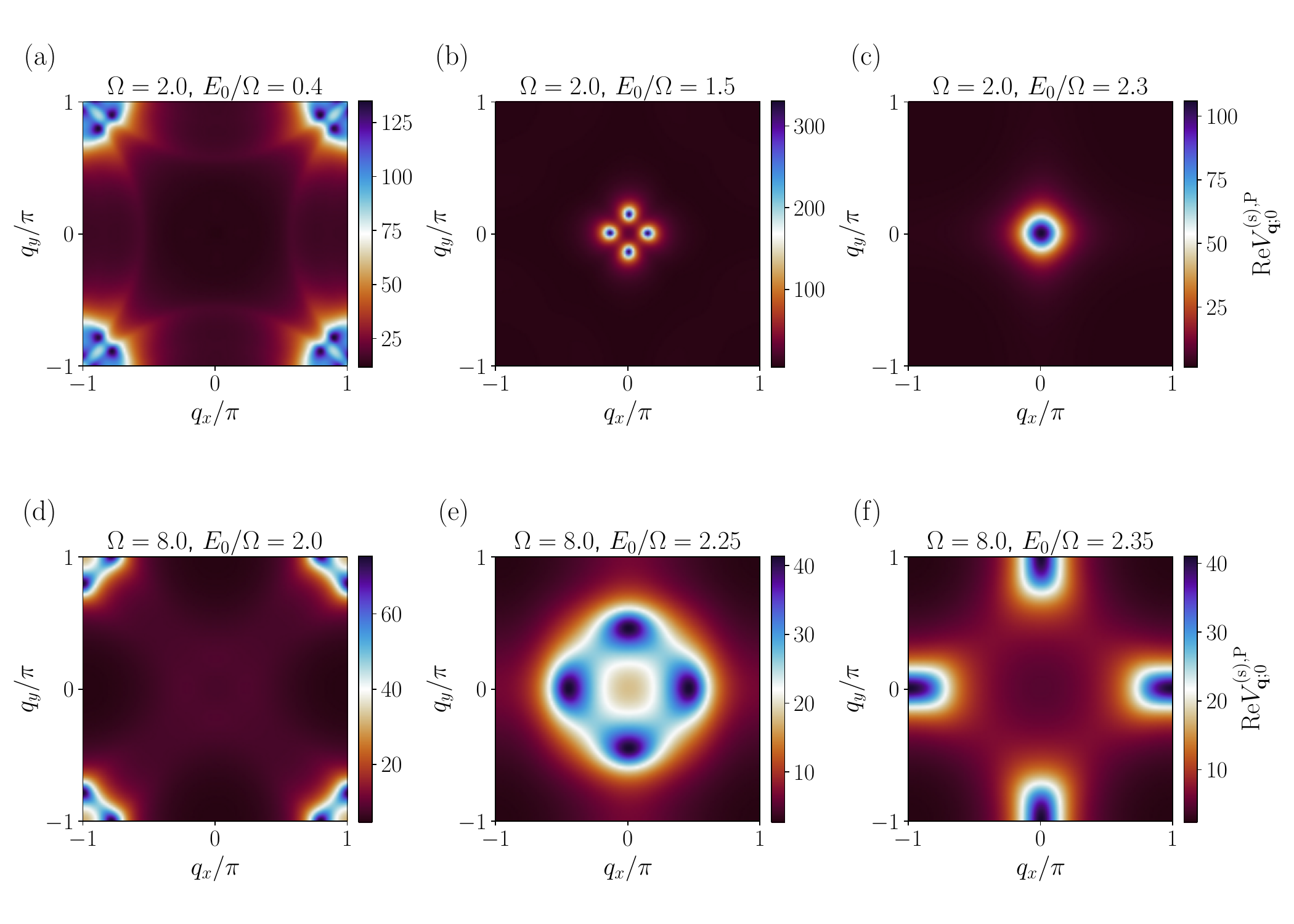}
\caption{The time-averaged effective pairing vertex in momentum space (spin-singlet) for representative drive parameters. Panel a) shows a sub-bandwidth drive protocol where the vertex is still peaked close to the $M$-point, while panels b) and c) show the vertex after the response transition and after the Floquet Lifshitz transition respectively. Panel d) shows the same for high frequency, where the peaks close to the $M$-point are enhanced. Panel e) shows the vertex close to the Floquet Lifshitz transition. Panel f) shows the vertex close to the dynamical localization point ($t_{\mathrm{eff}}/t'_{\mathrm{eff}}\ll1$), where the system is effectively an enlarged square lattice tilted by $\pi/4$, thus, the peaks sit at the $X$-point.}
\label{fig:pairing_vertex}
\end{figure}
The drive affects tunes the pairing vertex such that it can either enhance or suppress superconductivity by turning the nearest neighbor or next nearest neighbor components in real space attractive. In Fig.~\ref{fig:pairing_vertex} one can see the momentum resolved time-averaged pairing vertex for representative drive parameters. Peaks at the respective high-symmetry points reveal the expected ordering tendencies. Low amplitude driving keeps the peak close to the $M$-point, which favors AFM ordering and $d$-wave superconductivity. Intermediate-frequency, high-amplitude driving leads to a peak close to the $\Gamma$-point, thus, we expect FM ordering and, close to the Floquet Lifshitz transition (Fig.~\ref{fig:pairing_vertex}e)), we find spin-triplet $p$-wave superconductivity to stabilize. In the high-amplitude, high-frequency case, where $t_{\mathrm{eff}}<t'_{\mathrm{eff}}$ the peak moves to the $X$-point (Fig.~\ref{fig:pairing_vertex}f)), where one finds AFM order on an enlarged unit cell of the next-to-nearest neighbors. Here, the $d$-wave superconducting order mainly has $d_{xy}$ symmetry.\\
\begin{figure}
\includegraphics[width=1\linewidth]{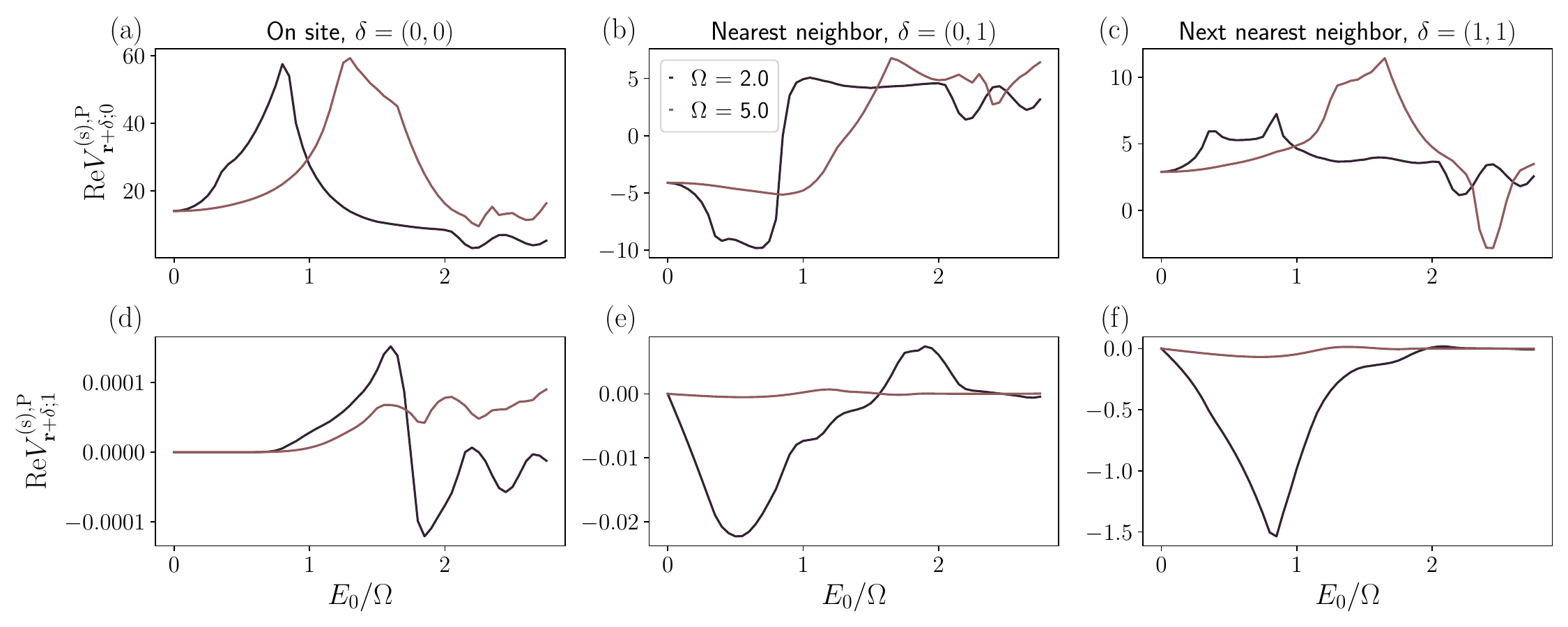}
\caption{The real space components of the pairing vertex as a function of drive amplitude for different drive frequencies. The lattice constant is set to $a=1$. The first row shows the $n=0$ component, i.e. the time-averaged vertex. The second row shows the $n=1$ component that describes oscillations of the vertex. (a) and (d) show the onsite interaction which is repulsive. (b) and (e) show the nearest neighbor interaction, which is attractive in the time average, facilitating $d_{x^2-y^2}$ order. (c) and (f) show the next-nearest neighbor interaction that remains repulsive in the time average but becomes attractive in the $n=1$ component. The magnitude of the $n=1$ terms is strongly suppressed by $\Omega$.}
\label{fig:pairing_vertex_real_space}
\end{figure}
To further understand the mechanism that generates the $id_{xy}$ symmetry components, it is insightful to investigate the real space structure of the effective pairing vertex, see Fig.~\ref{fig:pairing_vertex_real_space}. The bonds on which the effective interaction becomes attractive (negative) determine the gap symmetry. For $d_{xy}$-symmetry, an attractive interaction on the next-nearest neighbor bonds is required. The interaction on the next-nearest neighbor bonds only becomes attractive in the higher order $n=1$ components of the vertex. Therefore, oscillations in the effective interaction that are induced through the drive give the gap its $d+id$ character. This also explains the suppression of the $id_{xy}$ components, as these higher order terms are suppressed by $\Omega$.

\section{Symmetry of the superconducting phase}
\begin{figure}
\includegraphics[width=1\linewidth]{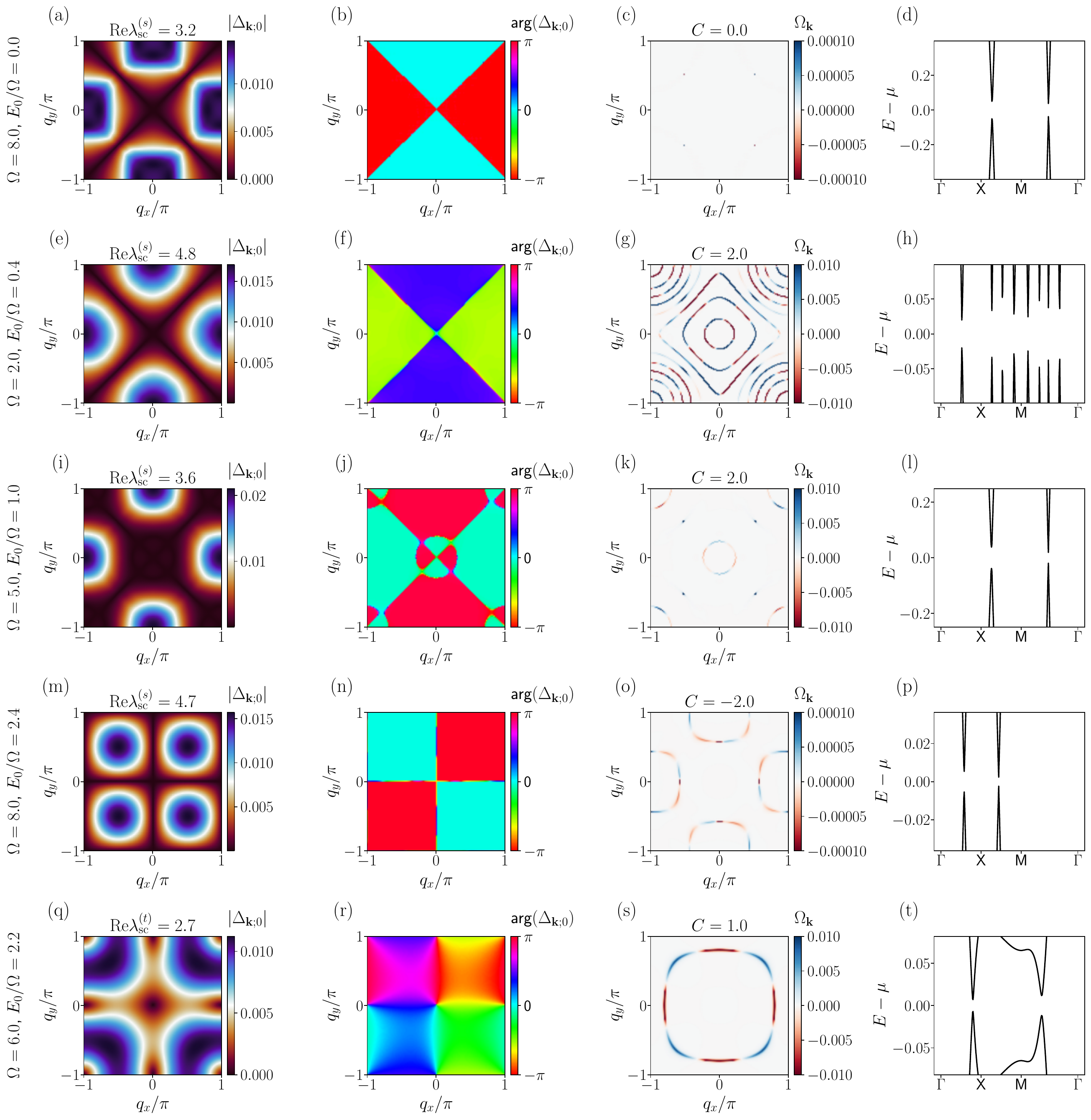}
\caption{The time-averaged superconducting singlet/triplet gaps for representative drive parameters. The columns show from left to right: 1) the absolute value of the time-averaged superconducting gap as obtained from the linearized gap equation, together with the real part of the corresponding (largest) eigenvalue, 2) the phase of the superconducting gap, 3) the Berry curvature computed with Fukui's method \cite{Fukui_2005} after plugging the gap into the Floquet BdG-Hamiltonian, 4) the Floquet bands around $\mu$ of the gapped BdG model. The rows show from top to bottom: i) the equilibrium case with topologically trivial $d_{x^2-y^2}$-symmetry ii) low frequency, low amplitude driving with chiral $d+id$-symmetry,  iii) intermediate frequency, low amplitude driving with chiral $d+id$-symmetry, iv) high frequency (at bandwidth $\Omega=W$), high amplitude driving at the dynamical localization point, showing the second kind of chiral $d+id$-symmetry with $d_{xy}$ being dominant, v) high frequency, high amplitude driving (just before the Floqet Lifshitz transition) showing chiral spin triplet $p+ip$-wave superconductivity. }
\label{fig:sc_gap}
\end{figure}

 The momentum structure of the time-averaged superconducting gap is shown in Fig.~\ref{fig:sc_gap} for $U=0.98U_c$ as obtained from the linearized gap equation. The topology of the gap is analyzed by plugging it into the BdG-Hamiltonian in Eq.~\eqref{eq: H_BdG} and using Fukui's method \cite{Fukui_2005}. In equilibrium, the symmetry of the superconducting order parameter in the square-lattice Hubbard model is known to be $d_{x^2 - y^2}$ due to the antiferromagnetic ordering tendency (Fig.~\ref{fig:sc_gap}(a)-(d)). Under driving, due to the time-reversal-symmetry breaking, the gap picks up a non-trivial complex phase that also generally leads to a complex eigenvalue. In the low-amplitude regime this leads to a moderate mixing of $id_{xy}$ symmetry regardless of drive frequency. Effectively, we find chiral $d+id$-wave superconductivity with a Chern number $C=2$ in the low-amplitude case, with an increased eigenvalue (Fig.~\ref{fig:sc_gap}(e)-(l)). We thus find agreement with the chiral superconductivity that has been reported in the strongly interacting limit~\cite{Anan_2024}.
Additionally, in the intermediate-to-high-frequency case (up to bandwidth $W$), we find intricate changes in ordering symmetry around the Floquet Lifshitz transition ($\tilde{t}_{\mathrm{eff}}/\tilde{t}'_{\mathrm{eff}}=1$). Triplet superconducting order is stabilized close to $U_c$ as AFM order is replaced by FM order (Fig.~\ref{fig:sc_gap}(q)-(t)). It has $p_x + ip_y$ symmetry with nontrivial topology ($C=1$). Beyond the Floquet Lifshitz transition close to the dynamical localization point ($\tilde{t}_{\mathrm{eff}}\to0)$) this changes into $d+id$-order again, but now the dominant component is the $d_{xy}$ symmetry (Fig.~\ref{fig:sc_gap}(m)-(p)) with Chern number $C=-2$.

\section{Floquet amplitude expansion}
In this section, we describe the derivation of the expansion results presented in Fig.~2(a) of the main text. 
We found that the qualitative behavior of $U_c$ at intermediate- to high-frequency and low-amplitude driving ($E_0/\Omega<1$) can be captured using an analytically tractable Floquet cutoff of $n_{\text{max}}=1$. Furthermore, we found that the overall feature of $U_c$ can be well reproduced by approximating it using the time-averaged components of the particle-hole susceptibility,

\begin{align}
    \chi^{\mathrm{ph};\mathrm{R}}_{\v{q};00}(\omega)  = \frac{i}{N} \sum_{\v{k},l} \int_{-\infty}^{\infty}&\frac{\d \omega'}{2\pi} G^{\mathrm{R}}_{\v{k};0,l}(\omega')G^{<}_{\v{k} - \v{q};l,0}(\omega'-\omega) + G^{<}_{\v{k};0,l}(\omega')G^{\mathrm{A}}_{\v{k} - \v{q};l,0}(\omega'-\omega)\;, \label{eq:ph susceptibility_TA}
\end{align}
and evaluating
\begin{align}
U_c \sim 1/\chi^{\mathrm{ph};\mathrm{R}}_{\v{q}_C;00}(0).    
\end{align}
To further simplify the analysis, we approximate $\chi^{\mathrm{ph};\mathrm{R}}_{\v{q}_C;00}(0)$ in the regime of low amplitude. Specifically, we expand $G^{R/A/<}_{00,\bm{k}}$,  up to second order of $\epsilon_{\bm{k},1}$, which corresponds to the off-diagonal element of Eq. (16) of the main text. This term is approximately proportional to $J_1(\frac{E_0}{\Omega})$ (the next-nearest neighbor hopping term is proportional to $J_1(\frac{ \sqrt{2} E}{\Omega})$, but since $t>>t^{\prime}$, its contribution is much smaller). In the low $\frac{E_0}{\Omega}$ limit, one has $J_0(\frac{E_0}{\Omega})>>J_1(\frac{E_0}{\Omega})$. The lowest order contribution corresponds to the Magnus limit, which reproduces the equilibrium results (including reservoir coupling) with the Magnus renormalized hopping amplitude, $t_{\mathrm{eff}} \rightarrow t\mathcal{J}_0(E_0/\Omega)$ and $t'_{\mathrm{eff}} \rightarrow t'\mathcal{J}_0(\sqrt{2}E_0/\Omega)$. The next lowest order (second order) correction is shown in Fig.~2(a) in the main text. Using the following notations, 
\begin{align}
   d_{n,\bm{k}}\left(\omega \right) &=(\omega+ n \Omega +i \gamma - \epsilon_{\bm{k},0})^{-1}, \\
   x_{\bm{k}} &=  |\epsilon_{\bm{k}, 1}|^2, \\
   R_n \left(\omega \right)&=  2 i \gamma f\left( \omega+ n \Omega \right),
\end{align}
the expansion up to the second order in $\epsilon_{\bm{k}, 1}$ (i.e., first order in $x$) is obtained as
\begin{align}
    G^{R}_{00}& \sim d_{0} + |\epsilon_{\bm{k}, 1}|^2 d_{0} \left(d_{-1}+d_{1}  \right) d_{0}^{-1} \label{eq:GR_tridigonal_expand_Ew}\\
     G^{<}_{00}  &\sim R_0 \left|  d_0 \right|^2 + x \left\{ R_0 2\text{Re}    \left[d_{0}  d_{0}^{\ast} \left(d_{-1}^{\ast}+d_{1}^{\ast}  \right) d_{0}^{\ast}\right] + R_{-1} \left|  d_0 d_{-1}   \right|^2+   R_{1} \left|  d_0 d_{1}   \right|^2  \right\},
\end{align}
where the momentum indices $\bm{k}$ and frequency dependence $\omega$ have been omitted for brevity. Applying the same procedure to $G^{R/A/<}_{0\pm1,\bm{k}}$ and $G^{R/A/<}_{\pm1 0,\bm{k}}$, i.e. expanding each Green's function up to the second order of $\epsilon_{\bm{k},1}$, and performing straightforward algebraic manipulations, we obtain $\chi^{\mathrm{ph};\mathrm{R}}_{\v{q};00}(\omega)$ as follows:
\begin{align}
\chi^{\mathrm{ph};\mathrm{R}}_{\v{q};00}(\omega) &=\chi^{\mathrm{ph};\mathrm{R}}_{\text{mag},\v{q}}(\omega)  +  \delta \chi^{\mathrm{ph};\mathrm{R}}_{\v{q}}(\omega)  \label{eq:PiR_sep}\\
   \delta \chi^{\mathrm{ph};\mathrm{R}}_{\v{q}}(\omega)  &\equiv \frac{\mathrm{i}}{2 \pi  N} \sum_{\bm{k}}\left( \left[|\epsilon_{\bm{k},1}|^2  d\chi^{\mathrm{ph};\mathrm{R}}_{\bm{k},\bm{q},0}(\omega) -|\epsilon_{\bm{k}-\bm{q},1}|^2  d\chi^{\mathrm{ph};\mathrm{R},\ast}_{\bm{k}-\bm{q},\bm{q},0}(-\omega) \right] \right. \notag\\
&+ \epsilon_{\bm{k},1} \epsilon_{\bm{k}-\bm{q},1}\left[  d\chi^{\mathrm{ph};\mathrm{R}}_{\bm{k},\bm{q},1}(\omega)    +d\chi^{\mathrm{ph};\mathrm{R},\ast}_{\bm{k}-\bm{q},\bm{q},1}(-\omega)  \right] \notag\\
&+\left.  \epsilon_{\bm{k},1} \epsilon_{\bm{k}-\bm{q},1}^{\ast}\left[ d\chi^{\mathrm{ph};\mathrm{R}}_{\bm{k},\bm{q},2}(\omega)    +d\chi^{\mathrm{ph};\mathrm{R},\ast}_{\bm{k}-\bm{q},\bm{q},2}(-\omega)     \right] \right)\\
   d\chi^{\mathrm{ph};\mathrm{R}}_{\bm{k},\bm{q},n}(\omega)& \equiv \int d\omega'  M_{\bm{k q},n}(\omega', \omega)   \\
   M_{\bm{k q},0}(\omega', \omega)&\equiv\left[ 
   R_{0}\left(\omega^{\prime}-\omega\right) d_{0,\bm{k}} (d_{-1,\bm{k}}+d_{1,\bm{k}})d_{0,\bm{k}}d_{0,\bm{k}-\bm{q}}d_{0,\bm{k}-\bm{q}}^{\ast} \right. \notag\\   
  &+ R_{0}\left(\omega\right) d_{0,\bm{k}} (d_{-1,\bm{k}}+d_{1,\bm{k}})d_{0,\bm{k}}d_{0,\bm{k}}^{\ast} d_{0,\bm{k}-\bm{q}}^{\ast} \notag\\
  &+ R_{0}\left(\omega\right) d_{0,\bm{k}} d_{0,\bm{k}}^{\ast} (d_{-1,\bm{k}}^{\ast}+d_{1,\bm{k}}^{\ast})d_{0,\bm{k}}^{\ast} d_{0,\bm{k}-\bm{q}}^{\ast} \notag\\
  &+R_{-1}\left(\omega\right) d_{0,\bm{k}} (d_{-1,\bm{k}}d_{-1,\bm{k}}^{\ast}) d_{0,\bm{k}}^{\ast} d_{0,\bm{k}-\bm{q}}^{\ast} \notag\\
  &+\left.R_{1}\left(\omega\right) d_{0,\bm{k}} (d_{1,\bm{k}}d_{1,\bm{k}}^{\ast}) d_{0,\bm{k}}^{\ast} d_{0,\bm{k}-\bm{q}}^{\ast}    \right]\\
  M_{\bm{k q},1}(\omega', \omega)&\equiv\left[R_{-1}\left(\omega'-\omega\right)  d_{0,\bm{k}} d_{-1,\bm{k}} d_{-1,\bm{k}-\bm{q}} d_{0,\bm{k}-\bm{q}}^{\ast} d_{-1,\bm{k}-\bm{q}}^{\ast}    \right. \notag\\   
  &+\left. R_{1}\left(\omega'-\omega\right)  d_{0,\bm{k}} d_{1,\bm{k}} d_{1,\bm{k}-\bm{q}} d_{0,\bm{k}-\bm{q}}^{\ast} d_{1,\bm{k}-\bm{q}}^{\ast}      \right]\\
  M_{\bm{k q},2}(\omega', \omega)&\equiv\left[R_{2}\left(\omega'-\omega\right)   d_{0,\bm{k}} d_{-1,\bm{k}}    d_{-1,\bm{k}-\bm{q}} d_{0,\bm{k}-\bm{q}} d_{0,\bm{k}-\bm{q}}^{\ast}       \right. \notag\\   
  &+\left. R_{2}\left(\omega'-\omega\right)   d_{0,\bm{k}} d_{1,\bm{k}}    d_{1,\bm{k}-\bm{q}} d_{0,\bm{k}-\bm{q}} d_{0,\bm{k}-\bm{q}}^{\ast}           \right]
\end{align}
where we suppressed the frequency dependence $\omega$ of $d_{n,\bm{k}}$ by using implicit notations:
\begin{align}
 d_{n,\bm{k}}&\rightarrow d_{n,\bm{k}}\left(\omega\right)   \\
 d_{n,\bm{k}-\bm{q}}&\rightarrow d_{n,\bm{k}-\bm{q}}\left(\omega^{\prime}-\omega\right).   
\end{align}
In \Cref{eq:PiR_sep}$, \chi^{\mathrm{ph};\mathrm{R}}_{\text{mag},\v{q}}(\omega)$ denotes the contribution of the Magnus limit discussed above, where the hopping amplitudes are renormalized  ($t_{\mathrm{eff}} \rightarrow t\mathcal{J}_0(E_0/\Omega)$ and $t'_{\mathrm{eff}} \rightarrow t'\mathcal{J}_0(\sqrt{2}E_0/\Omega)$). This component corresponds to the black line in Fig.~2(a) in the main text (with $\Omega=8$). 

We evaluate \Cref{eq:PiR_sep} for $\omega=0$ and $\v{q}=\v{q}_C$, where $\v{q}_C$ is determined from the exact calculation described in the main text. According to our analysis, $\v{q}_C$ can also be accurately obtained from \Cref{eq:PiR_sep} up to $E_0/\Omega \leq 0.4$ for $\Omega=2$, $E_0/\Omega \leq 0.8$ for $\Omega=4$, $E_0/\Omega \leq 2$ for $\Omega=6$, and $E_0/\Omega \leq 2.2$ for $\Omega=8$.

\end{document}